\documentclass[11pt]{article}

\title{Einstein's Photon Concept Quantified\\ by the Bohr Model of the Photon}

\author{Geoffrey Hunter\thanks{Correspondent: ghunter@yorku.ca, Chemistry Department, York University.}, 
Marian Kowalski\thanks{now with Optech Inc., 100 Wildcat Rd, Toronto, Canada.}
and Camil Alexandrescu\thanks{Physics Department, York University.},\\\phantom{line}
\\
{\it York University, Toronto, Ontario, Canada M3J$\,$1P3}}

\begin{document}

\maketitle

\begin{abstract}
The photon is modeled as a monochromatic solution of Maxwell's equations confined as
a soliton wave by the  principle of causality of special relativity.
The soliton travels rectilinearly at the speed of light.  
The solution can represent any 
of the known polarization (spin) states of the photon.  
For circularly polarized states the soliton's envelope is a circular ellipsoid 
 whose length 
is the observed wavelength ($\lambda$), and whose diameter is $\lambda/\pi$;
this envelope contains the electromagnetic energy of the wave ($h\nu=hc/\lambda$).
The predicted size and shape is confirmed by experimental measurements: 
of the sub-picosecond time delay of the photo-electric effect, of the
attenuation of undiffracted transmission through slits narrower than the 
soliton's diameter of $\lambda/\pi$, and by the threshold
intensity required for the onset of multiphoton absorption in focussed
laser beams.  Inside the envelope the wave's amplitude increases linearly 
with the radial distance from the axis of propagation, being zero on the axis.
Outside the envelope the wave is evanescent with an amplitude that decreases
inversely with the radial distance from the axis.
The evanescent wave is responsible for the observed double-slit interference phenomenon.
\end{abstract}


\section{Einstein's Concept of the Photon}

In 1905 Einstein published a celebrated paper popularly known as 
``the photo-electric paper'', for which he was awarded the Nobel prize
some 16 years later \cite{Einstein}.  While this paper explains other physical
phenomena in addition to the photo-electric effect, the unifying concept is that:
\begin{quote}
``The energy in a beam of light is not uniformly distributed as in a classical plane
wave, but is localized in packets of electromagnetic radiation, each packet having
an energy $h\nu = hc/\lambda$, where $h$ is Planck's constant, and $\nu$ and $\lambda$
are the frequency and wavelength of the radiation.''
\end{quote}
Einstein called his packets, ``light-quanta''; the modern term, ``photon'' was coined 
by G.N.Lewis \cite{Lewis}.
This localized-packet concept explains the photo-electric effect by a quasi-chemical
equation:
\begin{eqnarray}\label{PEE}
{\rm photon}\; +\; {\rm atom}\; \rightarrow\; {\rm emitted\_electron}\; +\; {\rm positive\_ion}
\end{eqnarray}
the energy of the emitted electron being equal to the energy of the absorbed photon minus
the energy required to remove the electron from the atom within the surface of the solid
forming the photocell:
This equation (\ref{PEE}) predicts the experimentally observed 
characteristics of the photo-electric effect:
\begin{enumerate}
\item all the emitted electrons have the same kinetic energy when the light is monochromatic,
\item this kinetic energy increases as the frequency of the light increases (as the wavelength
decreases),
\item there is a minimum frequency ($\nu_0$) of the light (maximum wavelength) 
below which no electrons are emitted; the energy, $h\nu_0$ is the energy required
to remove an electron from the surface of the photocell (binding energy of the electron,
ionization energy of the atom,
within the solid surface),
\item the kinetic energy of the emitted electrons is $h\nu$$-$$h\nu_0$, which is observed
as the voltage of the photo-cell being given by:  $V = (h\nu$$-$$h\nu_0)/e$
where $e$ is the electric charge of the electron,\footnote{$\nu_0$ is measured as the 
reverse (stopping) voltage, $(h\nu_0)/e$, needed to just stop the flow of current
from the photocell.}
\item the electric current generated by the photocell (rate of charge flow $\equiv$ 
electrons per unit time) is
proportional to the radiation intensity absorbed by the cell
(equivalent to number of photons per unit area of cell surface absorbed in unit time).
\end{enumerate}

\subsection{The Photon in Quantum Field Theory and the Bohr Model}
Although the quantization of the radiation field in terms of photons of energy $h\nu$
 became part of the standard language of quantum optics \cite{MW,Zubairy},
Einstein's original concept of the photon as a localized packet of 
electromagnetic radiation was discarded with the ascendency of quantum 
mechanics in the mid-1920s.\footnote{A corollary of Einstein's localized packet concept
of the photon is that much of the cross-sectional area of a beam of light is
empty space, the proportion of the beam's area occupied by packets
increasing with increasing light intensity.}
Theories of the wave function of the photon \cite{Sipe,Bila} 
preclude this localization, and even the quantum theory of
the photo-electric effect models the light as a plane wave \cite[pp.215-224]{Evanwave1}.

In contrast, the Bohr model of the photon predicts the size and shape of photons, and is
thus a quantification of Einstein's localized packet concept.  This prediction of size and
shape was not an primary objective of the Bohr model of the photon 
\cite{HWphoton}; the size and shape resulted from imposition of the principle of causality
on the chosen solutions of Maxwell's equations.  The result was that a circularly polarized
photon is a monochromatic electromagnetic traveling wave confined within a circular 
ellipsoid of length equal to the wavelength ($\lambda$), and of diameter 
$\lambda/\!\pi$; i.e.~an egg-shaped solitary wave propagating along the long axis of the
ellipsoid.\footnote{The ellipsoid is 3 times (accurately $\pi$ times) as long as its diameter.}
This prediction of size and shape was most important because it provided a basis for
comparison with experimental observations.  The comparison produced agreement between the
predicted size and shape and those inferred from several experimental measurements and observations.  This agreement makes the Bohr model worthy of serious consideration
even though its theoretical basis (a quantized solution of Maxwell's equations confined by
causality) is dissimilar from the widely accepted quantum field 
theory of light \cite{MW,Zubairy}.

This ellipsoidal soliton model of the photon is a Bohr model in the sense 
that it is a solution of the classical equations of motion that is subsequently quantized.  
In Bohr's well-known model of the 
hydrogen atom the classical equations are Newton's equations for the motion of an electron 
within the field of a proton, whereas for the photon (light regarded as electromagnetic 
radiation) the appropriate classical equations are Maxwell's equations in vacuum.
In Bohr's model of the hydrogen atom the quantization makes the angular momentum of the 
electron an integer multiple of Planck's constant, $\hbar=h/2\pi$.  In the Bohr model of 
the photon the quantization of the photon's angular momentum arises from an appropriately 
chosen solution of Maxwell's equations: in addition, the energy of the oscillating 
electromagnetic field (integrated over the volume of the ellipsoid) is quantized to 
be $h\nu$ - the known energy of the photon; this quantization fixes the amplitude of the
wave, and is analogous to the imposed quantization of angular momentum in Bohr's model
of the hydrogen atom; the analogy extends to quantization of the energy being
$n\,h\nu$ with $n\!>\!1$ representing a multiphoton.\footnote{Observations indicate
that multiphotons have a strong tendency to separate laterally into single photons
moving along parallel propagation axes.  This instability and the stability of a photon
with just one $h\nu$ of energy is an outstanding mystery of physics, whose eventually
resolution should yield a profound insight into the nature of Planck's constant.}

  The full theoretical derivation of the Bohr model is presented in \cite{HWphoton}
together with supporting experimental evidence; here the theory and experimental support
 is summarized and augmented by recent ideas pertaining to how a
solitary wave can exhibit two-slit interference.

\subsection{The Bohr Model of the Photon Summarized}

The solution of Maxwell's equations was chosen to be a monochromatic traveling wave having the observed
angular momentum of the photon; i.e.~a spin of $\pm\hbar$; constant parameters multiplying each of these 
spin states allows for representation of all the known polarization states of light.

The chosen solution of Maxwell's equations is confined within a finite space-time region by the principle 
of Special Relativity that causally related events must be separated by time-like intervals.
With the idea that a photon is self-causing as it propagates, causality imposes the condition that 
events within the wave having the same phase must be separated by time-like intervals.  In the limit where
the interval becomes null (light-like), causality leads to the inference that the length of the photon along 
its axis of propagation is the wavelength, $\lambda$.\footnote{or equivalently in time, the period of oscillation
$\tau=\nu^{-1}$.}  In addition, for circularly polarized states the causally connected field is contained
within a circular ellipsoid with maximum diameter (transverse to the axis of propagation) of $\lambda\!/\!\pi$;
the length of the ellipsoid (along the axis of propagation) is the 
wavelength.\footnote{The ellipsoidal soliton
can be visualized as an egg, or as an american/rugby football.}

This modeling of the photon as an ellipsoidal soliton arises from the imposition of 
causality upon the solution of Maxwell's equations (which are linear and homogeneous) 
whereas non-relativistic solitons arise as solutions
of non-linear differential equations \cite{solitons}.

The size and shape of the soliton allowed for quantization of its energy; 
the wave's electromagnetic energy, ${\bf E}^2+{\bf H}^2$, integrated over the volume 
of the ellipsoid, was set to $h\nu$.\footnote{Or in general, to $n\,h\nu$, analogous
with Bohr's quantization of angular momentum as $n\,\hbar$.}  
This fixed the amplitude of the wave and led to an expression for the 
average intensity within the photon-soliton 
\cite[eqn.57]{HWphoton}:\footnote{The photon's intrinsic intensity is not uniform: 
being proportional to the radius ($r$) squared\\ 
it is zero on the axis of propagation and maximal 
at the ellipsoid's maximum radius of $r=\lambda/2\pi$.}
\begin{equation}\label{Ip}
I_p = \frac{4\pi h c^2}{\lambda^4}
\end{equation}

\subsection{Experimental Confirmation of the Soliton}

Experiment confirms the predicted size, shape and intrinsic intensity of
the photon:
\begin{itemize}
\item its length of $\lambda$ is confirmed by:
\begin{itemize}
\item the generation of laser pulses that are just a few periods long;
\item for the radiation from an atom to be monochromatic (as observed),\\ 
the emission must take place within one period, $\tau$,  \cite{Kowalski};
\item the sub-picosecond response time of the photoelectric effect 
\cite{Downer};\footnote{The predicted absorption time of the ellipsoidal photon
is its period of oscillation, $\tau=1/\nu$ = the transit time
of the ellipsoid past any point in space = the time to enter the surface of
the photocell: a few femtoseconds for visible light.}
\end{itemize}

\item the diameter of $\lambda\!/\!\pi$ is confirmed by:
\begin{itemize}
\item the attenuation of direct (undiffracted) transmission of circularly polarized 
light through slits narrower than $\lambda\!/\!\pi$: our own measurements of the
effective diameter of 
microwaves \cite[p.166]{HWphoton} 
confirmed this within the experimental error of 0.5\%;
\item the resolving power of a microscope (with monochromatic light) being
``a little less than a third of the wavelength'';  
$\lambda\!/\!\pi$ is 5\% less than $\lambda\!/3$, \cite{Starling};
\end{itemize}
\item The predicted intrinsic intensity (given by eqn.\ref{Ip}) is the threshold (minimum) 
intensity to which a
laser beam must be focussed in order to produce multiphoton absoption: two 
experiments confirming
this (one with 650$nm$ light \cite{Verma}, the other with $\lambda$=10.5$\mu m$)
are described in \cite[p.165]{HWphoton}.
\end{itemize}

\subsection{Solution of Maxwell's Equations: the Photon's Wave Function}

Maxwell's equations \cite{Fewkes} relate the first derivatives of the six components of the electromagnetic
field; they comprise eight partial differential equations which must be satisfied 
simultaneously.\footnote{The equations are linear and homogeneous with constant coefficients.}
The key to finding appropriate solutions, is to 
differentiate to produce second derivatives followed by elimination of common terms 
between the resulting equations to yield the result that each Cartesian component of the 
field ($E_x, E_y, E_z, H_x, H_y, H_z$) separately satisfies d'Alembert's wave equation 
\cite{Fewkes}.\footnote{This only pertains for the {\em Cartesian} components; 
it does not prevail for the spherical or cylindrical components.}

For a wave traveling parallel to the $z$-axis at the speed of light, $c$, the solution must be any function
of $z-ct$ \cite{Coulson}, and if this wave is monochromatic the 
functional form is simply:\footnote{$S(z\!-\!ct)$ is an eigenfunction of 
Schr\"odinger operators: momentum 
in the direction of propagation, $\hat{p}_z=\frac{\hbar}{i}\frac{\partial}{\partial z}$, with
eigenvalue $h/\lambda$, and energy, 
$\hat{E}=-\frac{\hbar}{i}\frac{\partial}{\partial t}$, 
with eigenvalue $h\,c/\lambda=h\nu$, the physically known values for the photon.}
\begin{eqnarray}\label{travelwave}
S(z-ct) = \exp\{2\pi i(z-ct)/\lambda\}
\end{eqnarray}
When this form is adopted as a factor of the solution, insertion into d'Alembert's equation causes
a complete separation of $z$ and $t$ from the transverse coordinates 
($x\!=\!r\cos\phi,\; y\!=\!r\sin\phi$),\footnote{The separation is complete:
there is no
separation constant between the $z,t$ and the $r,\phi$ differential equations.} plane
polar coordinates ($r,\phi$) being chosen in preference to the Cartesian coordinates ($x,y$) in view of the
axial symmetry of the direction of propagation.

Separation of the radius, $r$, from the polar angle, $\phi$, produces the two ordinary differential equations:
\begin{eqnarray}\label{rfyeqns}
\frac{1}{\Phi(\phi)}\frac{d^2\Phi(\phi)}{d\phi^2} = m^2 = -
\frac{1}{R(r)}\left\{\frac{d^2R(r)}{dr^2}+\frac{1}{r}\frac{dR(r)}{dr}\right\}
\end{eqnarray}
where $m^2$ is the real separation constant introduced to separate $r$ from $\phi$.

The simplest solution of eqn.(\ref{rfyeqns}) is the plane wave ($m^2=0$); 
i.e.~$R(r)$ and $\Phi(\phi)$ both being constants.\footnote{Plane waves are widely used 
in the quantum field theory of light \cite{MW,Zubairy,Heitler}.}  
However, this solution was rejected as unphysical because light is observed to 
travel along very narrow beams.\footnote{A plane wave has field components that
have the same value throughout any plane perpendicular to the axis of propagation, 
and thus it is completely non-localized, contrary to observation that
light moves along very narrow beams.}

The next simplest solution of eqns.(\ref{rfyeqns}) is for $m^2=1$: i.e.~a factor of $r$ or $1/r$, with an angular
factor of $\exp\{i(\phi)\}$ or $\exp\{-i(\phi)\}$.  
These angular factors are
eigenfunctions of the $z$-component of angular momentum, ${\bf L_z}=\frac{\hbar}{i}\frac{\partial}{\partial\phi}$, 
in Schr\"odinger quantum mechanics \cite[p.217]{McQuarrie}, 
the eigenvalues of $\pm\hbar$ being those observed for the spin angular momentum of the photon; 
thus these solutions for $m^2$=$1$ are appropriate for the wavefunction of the photon:
\begin{equation}\label{wavesoln}
\psi(r,\phi,z\!-\!ct) = (\alpha r + \beta/r)
\left(A\exp\{i\phi\}+B\exp\{\!-i\phi\}\right)
 \exp\{2\pi i(z\!-\!ct)/\lambda\}
\end{equation}
Having determined this (the form in eqn.\ref{wavesoln}) 
as the appropriate solution of d'Alembert's equation, each of the 6 field components
($E_x, E_y, E_z, H_x, H_y, H_z$) will have this form, 
the coefficients ($\alpha,\beta,A,B$) being different
in each component.  
The relationships between the coefficients of different components were determined by 
Maxwell's equations.  This produced the inferences:
\begin{eqnarray}\nonumber
E_z&=&H_z=0\quad\quad({\rm no\ field\ along\ the\ axis\ of\ propagation:\ a
\ transverse\ wave})\\\label{HWfield}
E_x&=& (\alpha r + \beta/r)
\left(A\exp\{i\phi\}+B\exp\{\!-i\phi\}\right)
 \exp\{2\pi i(z\!-\!ct)/\lambda\}= \mu_0 c H_y\\\nonumber
E_y&=& i(\alpha r - \beta/r)
\left(A\exp\{i\phi\}-B\exp\{\!-i\phi\}\right)
 \exp\{2\pi i(z\!-\!ct)/\lambda\}= -\mu_0 c H_x
\end{eqnarray}

Imposition of the causality condition led to the result that if $A$ or $B$ is zero,
then the field must be contained within a circular ellipsoid of length $\lambda$ and 
cross-sectional diameter $\lambda/\pi$ \cite[\S2.5]{HWphoton}.

Since Maxwell's equations are linear and homogeneous they do not determine 
the amplitude of the solutions; this was
 determined by integration of the energy of the wave, 
${\bf E}^2+{\bf H}^2$.\footnote{This is analogous with Bohr's quantization of 
the electron's angular momentum  in his model of the hydrogen atom.}
This led to the realization that the form $1/r$ would cause a divergent contribution 
to the energy
at $r=0$, while the form $r$ would cause a similar divergence as $r\rightarrow\infty$. 
Thus, in view of the causality
condition limiting the domain of the field to an ellipsoid along the axis of propagation, 
it was decided to
discard the $1/r$ form and retain the $r$ form in order to produce a finite 
integrated energy. 
This discarding of the $1/r$ term (i.e.~$\beta$=0 in eqn.\ref{HWfield}) 
was concordant with the need to
make the field an eigenfunction of $L_z$ \cite[\S2.6]{HWphoton}.

This normalization of the amplitude of the photon's field 
yielded:\footnote{In \cite{HWphoton}
the amplitude squared ($\alpha^2$=$S_0^2$ in \cite[eqn.47]{HWphoton}) 
was given as, $\alpha^2= 64 n h c \pi^4/(\epsilon_0\lambda^6)$, which corresponds to
integration over a cylinder (length $\lambda$ and diameter $\lambda/\pi$) rather 
than the ellipsoid;
the factor of 120 in eqn.(\ref{normamp}) is correct for integration over the ellipsoid;
the relation $A^2+B^2=1$ was imposed, with $\alpha^2$ determined
by the energy integral.}
\begin{eqnarray}\label{normamp}
&& A^2+B^2=1\quad\quad{\rm and}\quad\quad
 \alpha^2= 120 n h c \pi^4/(\epsilon_0\lambda^6)
\end{eqnarray}

\subsection{The Soliton's Evanescent Wave}

An evanescent wave outside the ellipsoid is necessary as an adjunct to the theory 
presented in \cite{HWphoton}, because while the relativistic principle of causality
confines the wave within the ellipsoid, the radial dependence of the wave within the 
soliton is simply $r$, which is a maximum at the surface of the ellipsoid; 
physically the wave cannot sharply cut-off to zero at this surface;
it  must smoothly decay towards zero outside the ellipsoid; an evanescent wave  
decays in this way \cite[pp.103-108]{Evanwave1}.

The radial dependence of the evanescent wave is  $1/r$; i.e.~the solution of
Maxwell's equations (eqn.\ref{HWfield}) with $\alpha\!=\!0$.  
The intensity of this wave decreases as $1/r^2$
as the distance, $r$, from the axis increases.

J.J.$\,$Thomson derived the same solution (eqn.\ref{HWfield})
of Maxwell's equations in 1924 \cite{JJThomson};
he noted that a radial dependence of $r$ is appropriate near $r=0$, with $1/r$ 
being appropriate as $r\rightarrow\infty$, but he didn't pursue his analysis as 
far as deducing an ellipsoidal soliton, with the wave having the $r$ form within the 
ellipsoid, and the $1/r$ form outside the ellipsoid.

The $r$ dependence within the ellipsoid and the $1/r$ dependence outside the ellipsoid, 
makes the $r$-derivative of
the wave discontinuous on the surface of the ellipsoid.  
While this may appear to be unphysical, it is 
the same discontinuity exhibited by the gravitational force due to the mass of the Earth: 
on the assumption of a uniform density, 
the gravitational force inside the Earth is proportional to the radius, $r$,  
whereas outside the Earth it decreases like $1/r^2$ \cite{EarthsGravity}.

\subsection{Characteristics of the Photon's Evanescent Wave}

The polar components of the evanescent field are given by eqns.(38) of \cite{HWphoton} for $\alpha\!=\!0$
and $\beta$ given by eqn.(\ref{normbeta}), which show that none of these components have any dependence
upon the polar angle $\phi$, and that $E_r$ and $H_\phi$ are real, while $H_r$ and $E_\phi$ are imaginary:
\begin{eqnarray}\label{polarcomps}
E_r = \frac{\beta}{r}\left[A+B\right] = \mu_0\,c\,H_\phi\quad\quad
E_\phi = - i\,\frac{\beta}{r}\left[A-B\right] = - \mu_0\,c\,H_r
\end{eqnarray}
Independence of the angle, $\phi$, means that the evanescent wave carries none of the
angular momentum of the photon,\footnote{Because the operator for the $z$-component of 
angular momentum is ${\bf L_z}=\frac{\hbar}{i}\frac{\partial}{\partial\phi}$.}
and hence none of its energy; it is a truly evanescent wave \cite[pp.105-108]{Evanwave1}.

\subsection{Matching the Soliton and Evanescent Waves}\label{matchwaves}

While the gradient of the wave  has a cusp at $r\!=\!\lambda/(2\pi)$, 
the amplitude must be continuous at $r\!=\!\lambda/(2\pi)$; equating of the 
soliton and evanescent wave
amplitudes at $r\!=\!\lambda/(2\pi)$ produces:
\begin{eqnarray}\label{matchem}
\alpha\,r = \beta/r\quad\quad \rm for\quad\quad r\!=\!\lambda/(2\pi)
\end{eqnarray}
and since $\alpha^2$ is given by eqn.(\ref{normamp}) it follows that:
\begin{eqnarray}\label{normbeta}
 \beta^2= [\lambda/(2\pi)]^4\times 120 n h c \pi^4/(\epsilon_0\lambda^6)
= 7.5\, n h c /(\epsilon_0\lambda^2)
\end{eqnarray}

\paragraph{Orthogonality of the Radial Gradients}
The radial gradient of the soliton wave is simply the normalization constant, $\alpha$, while
that of the evanescent wave is $-\beta/r^2$.  Thus at the cusp where the two waves join 
(at $r\!=\!\lambda/(2\pi)$) the ratio of these gradients is:
\begin{eqnarray}\label{gradratio}
 {\rm ratio\ of\ gradients} = - \frac{\beta}{\alpha\,r^2} = -1\quad{\rm at}\quad r\!=\!\lambda/(2\pi)
\end{eqnarray}
Thus where the soliton and evanescent waves meet (at $r\!=\!\lambda/(2\pi)$) they are orthogonal
to each other - independent of the wavelength, $\lambda$.

The above matching of the soliton and evanescent waves  was made at the 
soliton's maximum diameter of $\lambda/\pi$; 
this raises the question of their matching at values of $z$ other than $z\!=\!0$;
i.e.~at other points on the ellipse:
\begin{eqnarray}\nonumber
&&\left(2\pi\,r\right)^2 + \left(2z\right)^2 = \lambda^2\\\label{ellipse}
\quad{\rm i.e.\ when}\quad
r&=&\frac{1}{2\pi}\sqrt{(\lambda)^2 - (2z)^2}\quad{\rm for}\quad -\frac{\lambda}{2}<z<+\frac{\lambda}{2}
\end{eqnarray}
It might appear natural to apply the matching condition of eqn.({\ref{matchem}}) for all values of $r$
specified in eqn.(\ref{ellipse}) to produce:
\begin{eqnarray}\nonumber
 \beta^2&=& \left[\frac{1}{2\pi}\sqrt{(\lambda)^2 - (2z)^2}\right]^4\times 120 n h c \pi^4/(\epsilon_0\lambda^6)
\\\label{normbetaz}
&=& \left[(\lambda)^2 - (2z)^2\right]^2\times 7.5 n h c /(\epsilon_0\lambda^6)
\end{eqnarray}
This would have the effect of making the amplitude of the evanescent wave, $\beta$, 
smaller as $z$ changes from $z\!=\!0$ to 
$z=\pm\frac{\lambda}{2}$, with $\beta$ being zero at these limits 
(the ends of the ellipsoid).

However, this conjecture would make $\beta$ a function of $z$ (as in eqn.\ref{normbetaz}) rather than a constant,
and hence the evanescent field (eqn.\ref{HWfield} for $\alpha\!=\!0, \beta\!\ne\!0$) would no longer be a 
a solution of Maxwell's equations.
The resolution of this physical vs.~mathematical paradox may be found
within the framework of General Relativity, in which the photon's local energy produces a non-Lorentzian metric.

\subsection{Diffraction and Interference}

The evanescent wave is believed to be responsible for the phenomena of diffraction and interference.
As a photon-soliton passes close to the edge of, or through a slit in, a material obstacle placed within the
beam of light, the interaction between the electrons within the obstacle and the photon's evanescent wave
will cause its path to bend as it passes by, the angle of bending (diffraction) being dependent upon the impact
parameter of the soliton's axis with the edge or slit.

Double slit interference can be understood by the soliton itself 
(like the $C_{60}$ molecules in Zeilinger's
experiment \cite{buckyball}) going through one slit or the other, 
while its evanescent wave extends over
both slits.  
The evanescent wave is like a classical continuous wave in extending throughout all
space, and hence the interference minima and maxima will appear at the 
same positions as predicted by Huygen's
theory.  However, the soliton model predicts that:
\begin{itemize}
\item the individual photons will arrive at local positions in the detection plane, 
whereas the classical
continuous wave model predicts a uniformly visible interference pattern: 
that the former (rather than the latter) 
is actually observed supports the soliton model \cite{buckyball};
\item the visibility of the interference pattern\footnote{Visibility, $V$, is defined by: 
$V=\frac{I_{\rm max}-I_{\rm min}}{I_{\rm max}+I_{\rm min}}$, $I_{\rm max}$ and $I_{\rm min}$ being the measured
intensities at the interference maxima and minima respectively; it has the range: $0\!\le\!V\!\le\!1$.}
will decrease with slit separation 
(because the intensity of the evanescent wave decreases like $1/r^{2}$, $r$
being the distance from the soliton's axis of propagation), 
whereas the classical continuous wave model
predicts a visibility independent of slit separation.  
This seems not to have been investigated experimentally \cite{MW,Evanwave1,BW}.
\end{itemize}

A double-slit experiment by Alkon \cite{Alkon} exhibits the expected interference pattern even though
the individual photons are constrained to pass through one slit or the other by an opaque barrier extending from
the source (a laser) up to the mid-point between the slits.\footnote{Alkon's experiment is the experimental
proof that the continuous wave concept that 
``the photon goes through both slits and interferes with itself'' is not correct.}  
This experiment demonstrates that the
particle-like photon (the Bohr model soliton) passes through one slit or the other, and yet its passage 
through this slit (and the subsequent diffraction) is affected by the presence of the other slit; this effect
of the other open slit is evidence for the existence of the evanescent wave surrounding the 
soliton.\footnote{Interaction between the evanescent waves of collaterally moving photon-solitons
could be the cause of the very small (but finite) divergence of a laser beam \cite[p.6]{Softley}.}

A causal model of diffraction has been proposed by Gryzinski \cite{Gryzinski}; it is based upon the photon
being a particle-like (localized) electromagnetic wave that interacts with the array of positive atomic nuclei 
and negative electrons within a solid, as it passes:
\begin{itemize}
\item through a crystal (Bragg diffraction of X-rays), or
\item adjacent to an edge of a sheet of the solid (an edge of a slit).
\end{itemize}
Gryzinsky's model of diffraction does not 
specify  the size or shape of the soliton, but it quantitatively explains
both Bragg diffraction and double-slit interference; his concept of the latter is that while the localized photon goes
through one slit, its wave extends to the other slit.  His theory is concordant with the Bohr model's evanescent wave, 
specifically because his localized model involves the concept that 
``the photon's electric field decreases when distance [from its center] increases''.

Gryzinsky pertinently cites Zeilinger's observation that each photon manifests its particle (localized) nature
in each detection event: the distribution of detection events\footnote{attributed in the continuous wave model
 to the wave going through both slits and self-interfering} 
only becomes manifest after a large number ($\ge 10^4$) of detection
events have been recorded \cite{buckyball}; each photon detection is a localized event.

The evanescent wave explanation for diffraction and interference is not readily 
invoked  for the Mach-Zender type
of interferometer, because the two alternative paths for the photon are 
typically separated by distances
over which the evanescent wave's intensity would have become negligible; 
a small difference (of the order of the wavelength) between the lengths 
of the two paths determines the observed interference pattern.  This 
observation requires further theoretical explanation.



\end{document}